\documentclass[twocolumn,prl,aps,showpacs,superscriptaddress,floatfix]{revtex4-1}
\usepackage[latin1]{inputenc}
\usepackage{bm}
\usepackage[usenames]{color}
\usepackage{multirow}
\usepackage{amssymb}
\usepackage{amsbsy}
\usepackage{amsmath}
\usepackage{stmaryrd}
\usepackage{graphicx}
\usepackage{epsfig}
\usepackage{placeins}
\usepackage{ulem}
\makeatletter

\newcommand{\tn}{\textnormal}

\begin{document}

\title{Proposal for measuring the finite-temperature Drude weight of integrable systems}

\author{C. Karrasch} 
\affiliation{Dahlem Center for Complex Quantum Systems and Fachbereich Physik, Freie Universit\"at Berlin, 14195 Berlin, Germany}

\author{T. Prosen}
\affiliation{Physics Department, Faculty of Mathematics and Physics, University of Ljubljana, Ljubljana, Slovenia}

\author{F. Heidrich-Meisner}
\affiliation{Department of Physics and Arnold Sommerfeld Center for Theoretical Physics,
Ludwig-Maximilians-Universit\"at M\"unchen, 80333 M\"unchen, Germany}
\affiliation{Kavli Institute for Theoretical Physics, University of California, Santa Barbara CA 93106, USA}

\date{\today}

\begin{abstract}
Integrable models such as the spin-1/2 Heisenberg chain, the Lieb-Liniger  or the one-dimensional Hubbard model are known to avoid thermalization, which 
was also demonstrated in several quantum-quench experiments. Another dramatic consequence of integrability is the zero-frequency anomaly in
transport coefficients, which results in  ballistic finite-temperature transport, despite the presence of strong interactions.
While this aspect of nonergodic dynamics has been known for a long time, there has so far not
been any unambiguous experimental realization thereof. We make a concrete proposal for the  observation ballistic transport  
via local quantum quench experiments in fermionic quantum-gas microscopes. Such an experiment would also unveil the coexistence of ballistic and diffusive transport channels
in one and the same system and provide a means of measuring finite-temperature Drude weights. The connection between local quenches and linear-response functions
is established via time-dependent Einstein relations.
\end{abstract}

\maketitle

{\it Introduction.--}
Nonergodic dynamics in closed many-body quantum systems is one of the most actively investigated branches of 
nonequilibrium physics \cite{Polkovnikov2011,Eisert2015,Langen2015r,Gogolin2016}. The canonical examples  are either Bethe-ansatz integrable one-dimensional (1D)
models such as the spin-1/2 XXZ chain,  the Fermi-Hubbard model \cite{Essler-book}, hard-core bosons \cite{Cazalilla2011} or many-body localized systems \cite{Basko2006,Nandkishore2015,Altman2015}.
Integrable systems possess an extensive set
of local conserved quantities that can constrain the long-time behavior in the relaxation dynamics starting from
nonequilibrium initial conditions \cite{Caux2011} induced by, e.g., quantum quenches.
This leads to the failure of these systems to thermalize with respect to standard thermodynamic ensembles (for a  review, see \cite{Vidmar2016}), rooted  in the  violation of  the 
eigenstate thermalization hypothesis \cite{Deutsch1991, Srednicki1994,Prosen1999,Rigol2008}. 

Another prominent consequence of integrability in clean systems is the possibility of anomalous transport properties at finite temperatures and in the
linear-response regime as was shown in a seminal paper by Zotos, Naef, and Prelov\v sek \cite{Zotos1997}.  
Within the Kubo formalism, one  decomposes conductivities into a regular part and a zero-frequency
contribution with the Drude weight $D$
\begin{equation}\label{eq:sigma}
\mbox{Re}\, \sigma(\omega) = 2 \pi D(T)\delta(\omega) + \sigma_{\rm reg}(\omega)\,.
\end{equation}
The presence of a nonzero Drude weight $D(T) > 0$
signals ballistic transport, which, at finite temperatures $T$, is an unusual behavior in a many-body system without momentum conservation.
The Drude weight can be related to the long-time value of current autocorrelation functions $C(t)= \langle I(t) I(0) \rangle_\textnormal{eq}/L$ ($I$ being the extensive current operator and $\langle A \rangle_{\rm eq} = Z^{-1} {\rm tr} [A e^{-H/T}]$) \cite{Zotos1997} and therefore, $D>0$
 directly implies nonergodic behavior in that correlation function. The connection to integrability is usually drawn via the Mazur inequality
\begin{equation}
D \geq \frac{1}{2 T^{\nu}L } \sum_\alpha \frac{|\langle I Q_\alpha \rangle_{\rm eq}|^2 }{\langle Q_\alpha^2\rangle_{\rm eq}}\,,
\label{eq:mazur}
\end{equation}
where $Q_\alpha$ are the local and pseudo-local \cite{qlreview} conserved charges for the Hamiltonian $H$ under consideration, i.e., $[H,Q_\alpha]=0$, $\langle Q^2_\alpha \rangle_{\rm eq}\propto L$, $\langle I Q_{\alpha}\rangle_{\rm eq}\propto L$, chosen such that $\langle Q_\alpha Q_{\alpha'}\rangle_{\rm eq} = 0$ for $\alpha\neq \alpha'$, and $\nu =1,2$ for charge(spin) and energy transport, respectively.

The most famous example is energy transport in the spin-1/2 XXZ chain, whose Hamiltonian is given by $H=\sum_l h_l$ with local terms
\begin{equation}\label{xxz}
h_l = J \left[ \frac{1}{2} (S^+_l S^-_{l+1} +\textnormal{h.c.}) + \Delta S^z_l S^z_{l+1} \right],
\end{equation}
where $S^\mu_l$, $\mu=x,y,z$ are the components of a spin-1/2 operator acting on site $l$ with $S^{\pm}_l = S^x_l \pm i S^y_l$.
$J$ is the exchange coupling and $\Delta$ denotes an exchange anisotropy.
In this model, 
   the energy-current operator $I_\textnormal{E}$ itself is conserved  $[H,I_\textnormal{E}]=0$ \cite{Zotos1997,Kluemper2002,Sakai2003}, 
rendering spin-1/2 XXZ chains ballistic thermal conductors.
 
Spin transport in the same model is also ballistic in its gapless phase $|\Delta| <1$ \cite{Zotos1996,Narozhny1998,HM2003,Herbrych2011,Karrasch2013,Karrasch2012,Karrasch2013,Zotos1999,Benz2005}.
The most debated case has been zero magnetization, since until the work \cite{Prosen2011} 
(see also \cite{Prosen2013,Mierzejewski2014,prosen2014NPB,qlreview,Pereira2014}) 
the relevant conservation laws were not known.
For $\Delta >1$, the common belief is that there is diffusion \cite{Steinigeweg2009,Znidaric2011,Steinigeweg2012,Karrasch2014} such that there is a  
oexistence of ballistic thermal  and diffusive spin transport.
 Notably, there is no final conclusion yet on the qualitative nature of spin transport at exactly  $\Delta =1$, i.e., the Heisenberg chain \cite{Alvarez2002,HM2003,Benz2005,Heidarian2007,Sirker2009,Sirker2011,Steinigeweg2014,Carmelo2015,Hild2014}, with superdiffusive dynamics being one possible scenario.

A similar situation applies to the  Fermi-Hubbard chain defined by
\begin{equation}\begin{split}\label{hub}
&h_l=-t_0  \sum_\sigma \left(c_{l\sigma}^\dagger c_{l+1\sigma}^{\phantom{\dagger}} + \tn{h.c.} \right) \\
&+ \frac{U}{2}\left[(n_{l\uparrow}-\frac{1}{2})(n_{l\downarrow}-\frac{1}{2})+(n_{l+1\uparrow}-\frac{1}{2})(n_{l+1\downarrow}-\frac{1}{2})\right],
\end{split}\end{equation}
where $c_{l\sigma}$ annihilates a fermion with spin $\sigma=\uparrow,\downarrow$ on site $l$, and $n_{l\sigma}=c_{l\sigma}^\dagger c_{l\sigma}^{\phantom{\dagger}}$, with $U$ the interaction strength and $t_0$ the hopping matrix element. 
The Hubbard model is  a ballistic thermal conductor at any $T>0$ \cite{Zotos1997,Karrasch2016}. 
At half filling, most studies indicate diffusive charge transport \cite{Carmelo2012,Prosen2012,Karrasch2014a,Jin2015}, while away from half filling, charge transport
is  ballistic. Thus, in the Hubbard chain there is a similar coexistence between a diffusive (charge) and a ballistic (energy) transport channel \cite{Karrasch2016}.

While the prediction of anomalous transport  in integrable models is very intriguing, its direct relevance 
to solid-state experiments is unclear since there, external scattering channels such as phonons or impurities will usually dominate the behavior \cite{shimshoni03,chernyshev05,rozhkov05a,boulat07,bartsch13,Chernyshev2016}. 
 Nevertheless, there has been an impressive series of experiments focussing mostly on thermal transport in quantum magnets \cite{Hess2001,Sologubenko2000,Sologubenko2000a,Sologubenko2001,Hess2003,Hofmann2003,Sologubenko2007,Hess2007,Hess2007a,Sologubenko2009,Otter2009,Otter2012,Montagnese2013,Hohensee2014}, reporting remarkably large thermal conductivities in 1D systems \cite{Hess2001,Hlubek2010}.
The cleanest evidence for ballistic transport in an integrable model has so far been observed in a strongly-interacting 1D Bose gas expanding in an optical lattice \cite{Ronzheimer2013}: the density profiles of these hard-core bosons 
are indistinguishable from noninteracting particles and the cloud thus expands ballistically.
An experimental demonstration of the aforementioned coexistence of a diffusive and a ballistic transport channel at finite temperatures  
thus remains open, as much as a quantitative measurement of the Drude weights, which so far were mostly considered rather academic quantities.

In our work, we propose a realistic set-up to measure finite-temperature Drude weights  using optical-lattice experiments with single-site resolution and addressing capabilities \cite{bakr09,sherson10,weitenberg11}.
The idea relies on preparing small local perturbations in spin- or charge densities and monitoring their spreading 
as a function of time. The time dependence of their width yields  information on whether transport is diffusive, ballistic, or something 
intermediate and access to diffusion constants {\it and} Drude weights can be gained from (generalized) Einstein relations. 
Einstein relations are usually quoted for the diffusion constant $\mathcal{D}= \sigma_{\rm dc}/\chi$ relating those to the dc conductivity $\sigma_{\rm dc}$ and
to the corresponding susceptibility $\chi$. The extension of such relations to ballistic transport {\it and} the transient regime before the asymptotic dynamics is reached is well-documented in the literature \cite{Steinigeweg2009a,Steinigeweg2009,yan2015} (albeit perhaps little appreciated beyond the diffusive case).

In this Letter we  demonstrate the validity of time-dependent generalized Einstein relations by a direct comparison of the time dependence of spatial variances in local quantum quenches to the expectation from time-dependent correlation functions for various types of transport in the spin-1/2 XXZ chain and the 1D Hubbard model, using
finite-temperature density matrix renormalization group simulations with the purification approach \cite{Verstraete2004p,Feiguin2005,Karrasch2012,Karrasch2013p}.
We  show that the time scales necessary to resolve ballistic dynamics are short enough to be observable in typical optical-lattice experiments. 
Moreover, the qualitative difference between the ballistic energy transport in both spin  and Hubbard chains versus the diffusive spin or charge transport (in the respective parameter regimes) can be unveiled on time scales  $\lesssim 10/J$ or $\lesssim 5/t_0$. 

\begin{figure}[t]
\includegraphics[width=0.49\columnwidth]{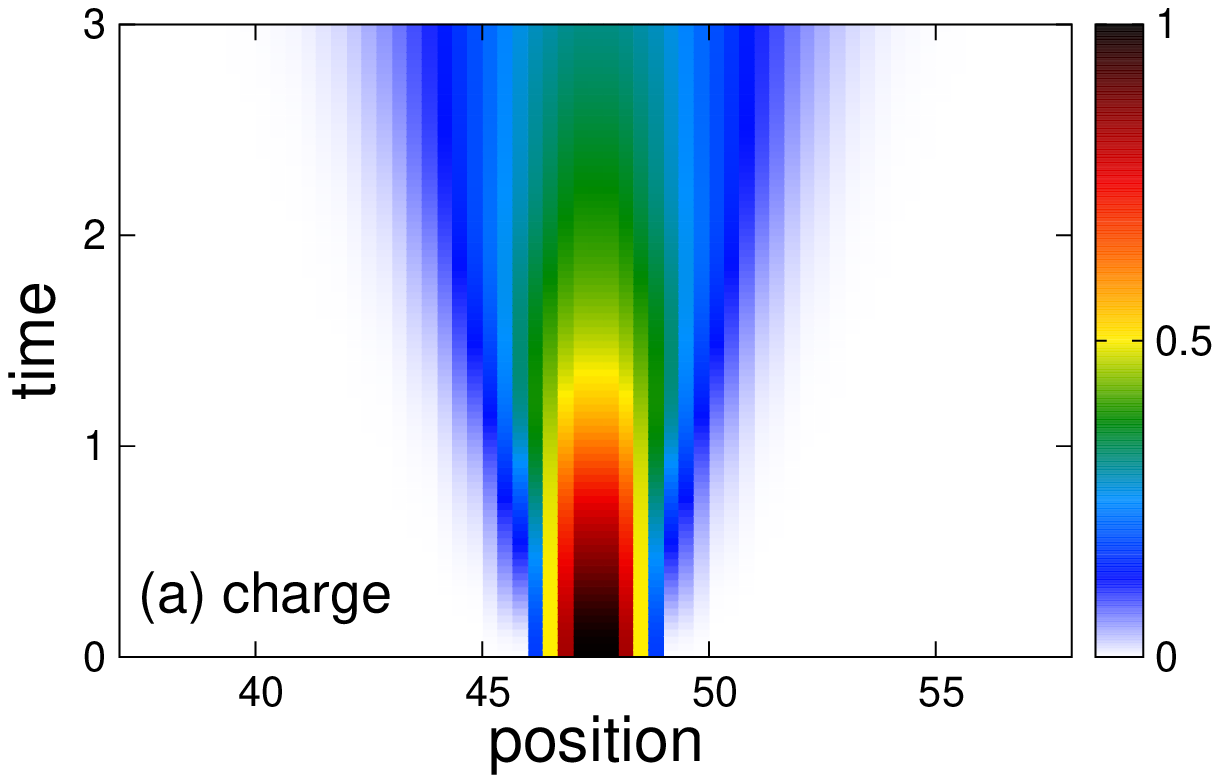}
\includegraphics[width=0.49\columnwidth]{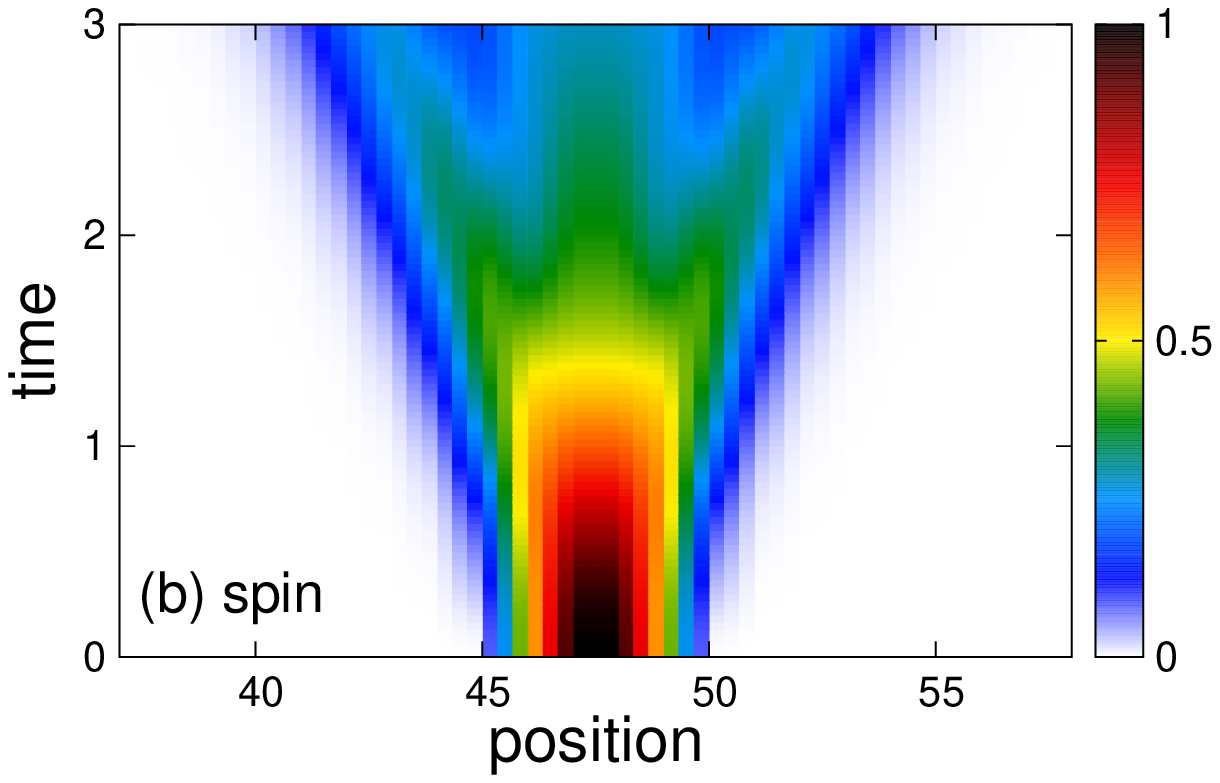}
\caption{(Color online) Example of numerical results for the spreading of  (a) a charge  and (b) a spin wave-packet induced by  a local quench in the half-filled 1D Fermi-Hubbard
model with $U/t_0=8$ and a spin-imbalance of  $\Delta n=(N_{\uparrow} - N_{\downarrow})/L=0.267$ at a temperature $T/t_0=20$,  where $N_\sigma$ is the number of  fermions
with spin $\sigma=\uparrow,\downarrow$. In (a) and (b), we show the charge and spin density differences $\langle n_{l\uparrow}\rangle+\langle n_{l\downarrow} \rangle-1$ and $\langle n_{l\uparrow}\rangle -\langle n_{l\downarrow} \rangle-\Delta n$ (both normalized to one), respectively.
In the initial state, we prepare two empty sites in the center of the system.
Our analysis (see the discussion of Fig.~\ref{Fig4})
suggests that
the charge dynamics is not ballistic, consistent with the prediction of diffusion \cite{Carmelo2012,Prosen2012,Karrasch2014,Karrasch2014a,ProsenPRE2014,Jin2015}, while spin dynamics is
ballistic.
\label{Fig1}
}
\end{figure}

The 1D Hubbard model provides the most promising avenue: several fermionic quantum-gas microscopes are now operational 
\cite{edge2015,Omran2015,haller2015,greif2016,boll2016,Cocchi2016,Cheuk2016,Parsons2015,Parsons2016}
and the 1D regime has  already been accessed \cite{boll2016}.
Unlike the ongoing quest to observe long-range antiferromagnetic correlations in this model \cite{schneider08,joerdens08,greif14,hart2015,boll2016,Parsons2016}, cooling to
low temperatures is not necessary since the ballistic transport properties are protected by conservation laws and are thus observable at any $T>0$. We propose to study the spin-imbalanced regime of a 1D Hubbard model at half filling, where charge transport is diffusive, but spin transport is ballistic. This effect is also of fundamental interest, being the first example of a coexistence of ballistic and diffusive finite temperature transport not involving thermal transport.

{\it Generalized time-dependent Einstein relations.--} The spatial variance of a density wave-packet can be related to linear-response functions via
 \cite{Steinigeweg2009a,Steinigeweg2009,yan2015} 
\begin{equation}\label{einstein}
\delta\sigma^2_\nu(t)=  \frac{2}{L\chi_\nu}\int_0^t dt_1 \int_0^{t_1} dt_2 \langle I_\nu(t_1)I_\nu(t_2)\rangle_\textnormal{eq} 
\end{equation}
where  $\delta \sigma^2_\nu(t)  =\sigma^2_\nu(t) -\sigma^2_\nu(0)$, $\nu={th,s,c}$ can be energy, spin, or charge (for definitions of $I_\nu$ and of $\sigma_{\nu}$,  
see \cite{SM}). At infinite temperature, the susceptibility $\chi_\nu$ is $ \chi_\textnormal{s} = 1/4$, $\chi_\textnormal{th}/J^2=1/8+\Delta^2/16$ for the XXZ chain and $ \chi_\textnormal{c} = 1/2$, $\chi_\textnormal{th}/t_0^2=1+U^2/16$ for the Hubbard model, but is also known
exactly at any finite temperature \cite{Kluemper2000,Essler-book}. A typical example for an 
initial state used in our simulations to induce the dynamics and the real-space and time-dependence of such as a  density perturbation is shown in Fig.~\ref{Fig1}.

In the ballistic case, $\langle I_\nu(t)I_\nu\rangle_\textnormal{eq}/L\to 2T^{\gamma}D_\nu$ (where $\gamma=2$ in the thermal case and $\gamma=1$ otherwise), and therefore, from Eq.~\eqref{einstein},
$\delta \sigma_\nu^2(t) \to \frac{T^\gamma D_\nu}{\chi_\nu } t^2\,$. 
For diffusive dynamics, one obtains 
$\delta \sigma_\nu^2(t) \to 2\mathcal{D}_\nu t$ with  $\mathcal{D}_\nu = \sigma_{\rm dc,\nu}/\chi_{\nu}$.

{\it Spin-1/2 XXZ chain.--}
We begin by discussing results for the spin-1/2 XXZ chain with $\Delta=0.5$ and $\Delta=1.5$, i.e., in its massless and massive regimes. Unless stated
otherwise, we consider infinite temperature $T=\infty$ (see also \cite{langer09,langer11,Karrasch2014,Steinigeweg2016}). 

\begin{figure}[t]
\includegraphics[width=0.9\columnwidth]{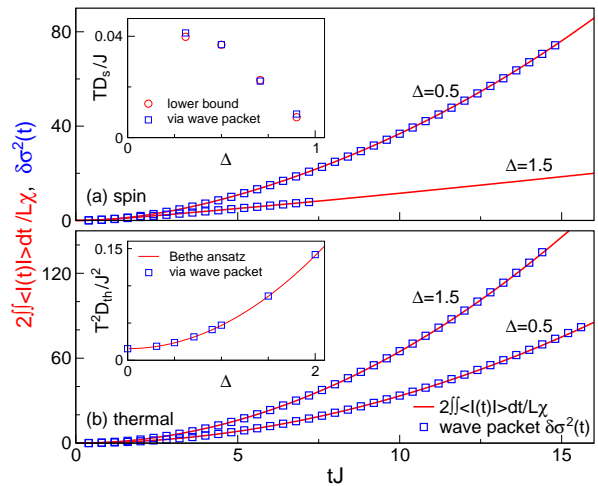}
\caption{(Color online) Time evolution of the square variance $\sigma_\nu^2(t)-\sigma_\nu^2(0)$ of a wave packet as well as of the r.h.s.~of Eq.~(\ref{einstein}) for the spin-1/2 XXZ spin chain at $T=\infty$ in  (a) a spin quench ($L_m=2$) and (b) an energy quench ($T_1/J\geq10$, $T_2=\infty$, $L_m=2$).
Insets: Comparison of the Drude weights extracted from the wave packets with the lower bound \cite{Prosen2013} (spin case in (a)) and with 
the Bethe-ansatz result \cite{Kluemper2002,Sakai2003} for $D_{th}$ in (b). The Drude weights $D_{s}$ and $D_{th}$ were extracted from fits to $a+bt+ct^2$ and $ct^2$ over the last third of curve, respectively.
}\label{Fig2}
\end{figure}

Here, we provide an example of the equivalence of the l.h.s.~and r.h.s.~of Eq.~\eqref{einstein} at any time $t>0$ for local quenches with a sufficiently
small amplitude (see \cite{SM} for an analysis of the influence of the amplitude). Figure~\ref{Fig2}(a) shows the results for a perturbation in the spin density 
that is initially prepared on $L_m =2$ sites (see \cite{SM} for definitions). For $\Delta=0.5$, 
there is a ballistic spreading with $\delta \sigma_s^2 \propto t^2$ while for $\Delta =1.5$ and at long times, the variance approaches a diffusive behavior 
with $\delta \sigma_s^2 \propto t$. The agreement between $\delta \sigma_s^2(t) $ (solid lines) and the Kubo formula (double-integral, r.h.s.~of Eq.~\eqref{einstein})
is excellent at all times.
While the diffusion constant was extracted already from such data for $\Delta>1$ \cite{Karrasch2014}, we here present results for the spin Drude weight for $\Delta<1$.
The comparison of wave-packet dynamics versus Kubo (double integral of current autocorrelations) show that either approach  gives
the same quantitative results. For the comparison with the literature we will use the method that gives access to the longest times and hence more accurate
estimates of the Drude weight. In the case of spin quenches, this is the time propagation of current correlations.
The results for $D_s$ computed for various values of $0<\Delta<1$ are shown in the inset of Fig.~\ref{Fig2}(a) and are in very good agreement with the
lower bound for $D_s$ from \cite{Prosen2013}, which  is believed to be exhaustive for the chosen values of $\Delta$.

Figure~\ref{Fig2}(b) shows the corresponding data for the thermal Drude weight, where the dynamics is induced by an energy quench,
 in which we embed a region of higher temperature $T_2$ into a background with $T_1<T_2$ 
(here, $T_1/J=10$ at $\Delta=0.5$, $T_1/J=20$ at $\Delta=1.5$, $T_2=\infty$, $L_m=2$ \cite{SM}). $D_{\rm th}$ is extracted from purely quadratic fits to $\delta \sigma_{th}(t)$.
The agreement with the exact Bethe-ansatz results \cite{Kluemper2002,Sakai2003} is excellent, as shown in the inset of Fig.~\ref{Fig2}(b). 

\begin{figure}[t]
\includegraphics[width=0.9\columnwidth]{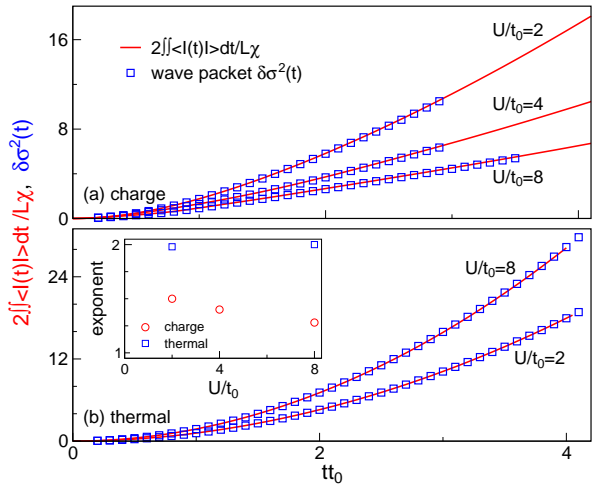}
\caption{(Color online) The same as in Fig.~\ref{Fig2}, but for the Fermi-Hubbard model at $T=\infty$ and considering   (a) a charge quench ($L_m=2$) 
and (b) an energy  quench ($T_1/t_0=20$, $T_2=\infty$, $L_m=2$). Inset: Exponents determined from a power-law fit 
to the last third of the curve.
}\label{Fig3}
\end{figure}

{\it 1D Hubbard model: half filling, zero magnetization.--}
Next we turn to the Fermi-Hubbard model with filling $n=1$ and a vanishing magnetization ($N_\uparrow = N_\downarrow$).
Figures~\ref{Fig3}(a)  and (b) show the comparison of the l.h.s.~and r.h.s.~of Eq.~\eqref{einstein} for a perturbation in the charge and an energy density, respectively, which agree well
in both cases. 
The difference in the qualitative behavior of charge and energy transport is the best unveiled by comparing the exponents of the variance at the longest times reached. The inset in Fig.~\ref{Fig3}(b) suggests that perturbations in the energy density spread ballistically as expected from theory \cite{Zotos1997,Karrasch2016} with an exponent of $2$,
while for the charge dynamics, the exponents are consistently smaller than 2 and the smaller the larger $U/t_0$ is.
We expect that the asymptotic diffusive behavior \cite{Karrasch2014a,Jin2015} will emerge at times longer than what is accessible in our simulations. 
Note that one can also study the time dependence of energy and charge density perturbations in quenches that touch both densities \cite{Karrasch2016}:
in that case, one can visualize the coexistence of ballistic energy transport with diffusive charge transport in the same time-evolution.

In principle, such an initial perturbation of both energy and charge density may thus yield the desired recipe for an experimental observation. 
While measuring double occupancy $d_l = \langle n_{l\uparrow} n_{l\downarrow}\rangle $, which yields the interaction energy,  
is standard (see, e.g., \cite{boll2016}), there is, unfortunately, as of now no good way of extracting energy densities in fermionic quantum-gas microscopes. We therefore turn to our most promising example, namely spin and charge dynamics at half filling but away from 
zero magnetization.

{\it 1D Hubbard model: half filling with spin imbalance.--}
We introduce a spin imbalance $\Delta n = (N_{\uparrow} - N_{\downarrow})/L$ while keeping the half-filling condition.
In this case, the spin transport becomes ballistic (different from the case $\Delta n=0$) since the 
spin current acquires a finite overlap with local conserved quantities \cite{Zotos1997}, constructed in \cite{shastry1986infinite}.
Charge transport, however, is expected to remain diffusive, since due to particle-hole symmetry and the half-filling condition the charge
current operator remains orthogonal to local conserved operators \cite{shastry1986infinite}. 
It would  in principle be possible that other quasi-local conserved operators existed in the Hubbard model which were not orthogonal to charge current in this case, similar as for the spin current in the massless regime of XXZ chain, but numerical computations \cite{MarcinUnpub} using the method of Ref. \cite{Mierzejewski2014} indicate that this is not the case.

\begin{figure}[t]
\includegraphics[width=0.9\columnwidth]{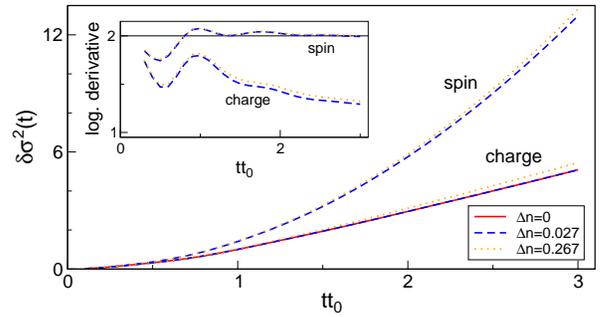}
\caption{(Color online)
(a) Comparison of $\sigma^2_{c}$ and  $\sigma^2_{s}$ for a charge quench ($L_m=2$) as a function of the imbalance at $U/t_0=8$ and $T/t_0=20$. Inset: Logarithmic derivative.}\label{Fig4}
\end{figure}

In order to illustrate the coexistence of ballistic spin and diffusive charge dynamics at $T>0$, 
we present results for  a charge quench with $T/t_0=20$ and $L_m=2$ empty sites, which for $\Delta n\not= 0$ also
affects the spin density.
Figure~\ref{Fig4} shows  the time dependence of the variances for different values of $\Delta n$,
while we present the exponents in the inset. 
First, we observe that spin propagates much faster than charge at $T=\infty$ and moreover, $\delta \sigma_s^2 \propto t^2$ as soon as
$\Delta n>0$, clearly indicating ballistic spin transport.
Second, the exponents for the charge dynamics are very close to unity for small $\Delta n$ and increase slightly as $\Delta n$ becomes large.
This behavior suggests diffusive charge dynamics, keeping in mind that the simulations may not yet have reached the asymptotic regime.
Note that this  difference in the charge versus spin dynamics is not spin-charge separation, which is a low-energy phenomenon with no qualitative difference in the
propagation of spin and charge.

We stress that  the validity of Einstein relations (\ref{einstein}) depends on the nature of the initial local perturbation.
Namely, the wave-packet of Fig.~\ref{Fig4} is formed by an experimentally realizable quench of producing  
$L_m=2$ empty sites $|00\rangle$ in the spin-imbalanced background. Unlike other initial states, 
it shows a curious discontinuity of the spin variance $\delta \sigma_s^2(t)$ at $T=\infty$, as for $\Delta n=0$ there is  no spin transport anymore and 
$\sigma^2_s(t)\equiv 0$, while at any finite $\Delta n>0$, spin propagates ballistically with a $\Delta n$-independent velocity.
 However, if one considers
 genuine local spin-quenches, e.g.,  preparing the central two sites in the state $|0\rangle |\!\!\uparrow\rangle + |\!\!\uparrow\rangle |0\rangle$, then $\sigma_s^2(t)$ nicely follows the linear-response prediction (\ref{einstein}) and smoothly switches from ballistic to diffusive growth as $\Delta n\to 0$ (results not shown). (For another recent study
of typical vs.~atypical initial states, see \cite{Steinigeweg2016}). We stress that in order to resolve the difference between ballistic spin and diffusive charge dynamics, it is not necessary to prepare
local quenches with a small amplitude. 

{\it Experimental aspects.--}
As we argued in the introduction, the  half-filled Hubbard chain away from zero magnetization provides  the most promising route to 
 experimentally observe the coexistence of a ballistic (here spin) and a diffusive transport channel (here charge) in one and the 
same model. 
A clear advantage of working directly with the Fermi-Hubbard chain is that one does not have to worry about
defects. For instance, in simulations of the Heisenberg model using a strongly interacting two-component Bose gas \cite{Fukuhara2013a},
holes or doublons invalidate the mapping to a pure spin model. Similarly, interactions have to be large for the mapping to be
valid, while the interaction strength in the Fermi-Hubbard chain plays no role for the qualitative aspects discussed here, yet may determine the
onset of asymptotic diffusive behavior.

Reaching low temperatures is not important as such to see the integrability-protected ballistic transport in either the spin-1/2 XXZ or in the 
Hubbard chain. 
However, most optical-lattice experiments have a harmonic trapping potential \cite{Bloch2008}. This breaks integrability as such, yet since we are working at
unit filling, a large region of the cloud will be in the Mott-insulating regime, with a constant density. This  requires that temperatures should ideally be low enough
to stabilize a sizable Mott plateau.

{\it Summary.--}
Experimentally measuring the finite-temperature Drude weight of an integrable 1D model has not been accomplished yet. We here proposed
to use local quenches, i.e., perturbations in spin-, charge- or energy density to induce a density wave-packet dynamics. 
We demonstrated that in ballistic regimes, the Drude weight can be extracted by simply monitoring the
time dependence of the spatial variance and measuring the prefactor. This relies on generalized time-dependent Einstein relations \cite{Steinigeweg2009a},
whose validity we verified in several examples including the spin-1/2 XXZ chain and Hubbard chains.
We identified the charge and spin dynamics in the half-filled Hubbard chain with a spin imbalance as the most promising candidate for 
an experiment, given that many ingredients are available in several groups, such as fermionic quantum microscopes \cite{edge2015,Omran2015,haller2015,greif2016,boll2016,Cocchi2016,Cheuk2016,Parsons2015,Parsons2016}, 1D Fermi-Hubbard systems \cite{boll2016}, and 
the ability to spin-selectively \cite{boll2016,Parsons2016} monitor the time-dependent spreading of local perturbations \cite{Fukuhara2013,Fukuhara2013a,Preiss2015}.
Such an experiment would visualize the coexistence of ballistic and diffusive transport channels in the same quantum model \cite{Zotos1997,langer11,Karrasch2014,Karrasch2016}, an extreme violation
of the standard behavior of Fermi liquids or Luttinger liquids.

{\it Acknowledgment.} 
We thank I. Bloch, C. Gross and R. Steinigeweg for very useful discussions and H. Spohn for comments on an earlier version of the manuscript.
F.H.-M. acknowledges the hospitality of KITP at UCSB, where  part of this research was carried out. 
This research was supported in part by the National Science Foundation under Grant No. NSF PHY11-25915. C.K. acknowledges support by the Emmy Noether program of the Deutsche Forschungsgemeinschaft (KA 3360/2-1). TP acknowledges support from Slovenian Research Agency grant N1-0025 and ERC grant OMNES.

% \newpage

\setcounter{figure}{0}
\setcounter{equation}{0}

\renewcommand{\thetable}{S\arabic{table}}
\renewcommand{\thefigure}{S\arabic{figure}}
\renewcommand{\theequation}{S\arabic{equation}}

\renewcommand{\thesection}{S\arabic{section}}

\bibliography{references}

\newpage
% \end{document}

\section*{Supplemental Material}

{\it Numerical method.--} We need to compute the real time evolution of linear-response correlation functions (current correlators)
\begin{equation}\label{timeeq}
\langle A(t)B\rangle_\tn{eq}  = Z^{-1}\tn{Tr}\big[ e^{-H/T} e^{iHt}Ae^{-iHt}B\big]
\end{equation}
and of observables in a nonequilibrium state (wave packets)
\begin{equation}\label{timenoneq}
\langle A(t)\rangle_0 = \tn{Tr}\big[ \rho_0 e^{iHt}Ae^{-iHt}\big]
\end{equation}
determined by $\rho_0$ with $[\rho_0,H]\neq0$. To this end, we employ the time-dependent \cite{vidal04,white04,daley04,schmitteckert04,vidal07} density matrix renormalization group method \cite{white92,schollwoeck05,schollwoeck11} in a matrix-product state \cite{fannes91,ostlund91,verstraete06,verstraete08} implementation. Finite temperatures \cite{Verstraete2004p,white09,barthel09,zwolak04,sirker05,barthel13} are incorporated via purification of the thermal density matrix. The real- and imaginary time evolution operators $e^{-iHt}$ and $e^{-H/T}$ are factorized by a fourth order Trotter-Suzuki decomposition. We keep the discarded weight during each individual `bond update' below a threshold value $\epsilon$. This leads to an exponential increase of the bond dimension $D$ during the real-time evolution. In order to access time scales as large as possible, we employ the finite-temperature disentangler introduced in Ref.~\onlinecite{Karrasch2012}, which uses the fact that purification is not unique to slow down the growth of $D$. Moreover, when computing correlation functions, we `exploit time translation invariance' \cite{barthel13},  rewrite $\langle A(t)B\rangle_\tn{eq}=\langle A(t/2)B(-t/2)\rangle_\tn{eq}$, and carry out two independent calculations for $A(t/2)$ as well as $B(-t/2)$. A similar trick can in principle be implemented when calculating the out-of-equilibrium $\langle A(t)\rangle_0$ \cite{kennes16}. Our calculations are performed 
using a system size of the order of $L\sim100$ sites. By comparing to other values of $L$ we have ensured that $L$ 
is large enough for the results to be effectively in the thermodynamic limit \cite{Karrasch2013}.

{\it Current correlations.--}
The spin current $I_\textnormal{s}$ in the XXZ chain and the charge current $I_\textnormal{c}$ in the Hubbard model take the standard form
\begin{equation}
I_\textnormal{s} = -\frac{iJ}{2}\sum_l S^+_l S^-_{l+1} + \textnormal{h.c.},~~
I_\textnormal{c}=-it_0 \sum_{l\sigma}c^\dagger_{l\sigma}c^{\phantom{\dagger}}_{l+1\sigma}+\tn{h.c.}~.
\end{equation}
The energy current is defined via a continuity equation, leading to $I_\tn{E} = i \sum_l [h_l,h_{l+1}]$.

\begin{figure}[t]
\includegraphics[width=0.9\columnwidth]{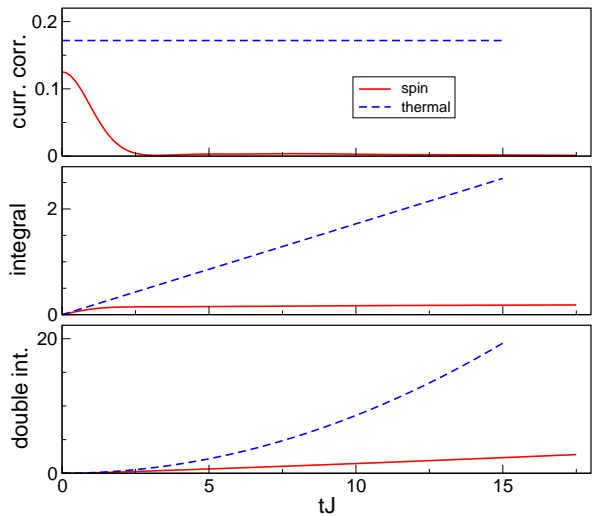}
\caption{(Color online)
(a) Current correlator and its integral and double-integral for ballistic (energy) and diffusive (spin) case $\Delta=1.5$ and $T=\infty$
for data from Fig.~\ref{Fig2}.}
\label{FIGS1}
\end{figure}

As outlined in the previous section, the linear-response current correlators $\langle I(t)I\rangle_\textnormal{eq}$ can be computed directly using the DMRG. Exemplary results are shown in Fig.~\ref{FIGS1} for the XXZ chain. In order to test the generalized Einstein relations, one need to determine the double integral of $\langle I(t)I\rangle_\textnormal{eq}$, which is straightforward to do numerically (see Fig.~\ref{FIGS1}).

\begin{figure}[t]
\includegraphics[width=0.9\columnwidth]{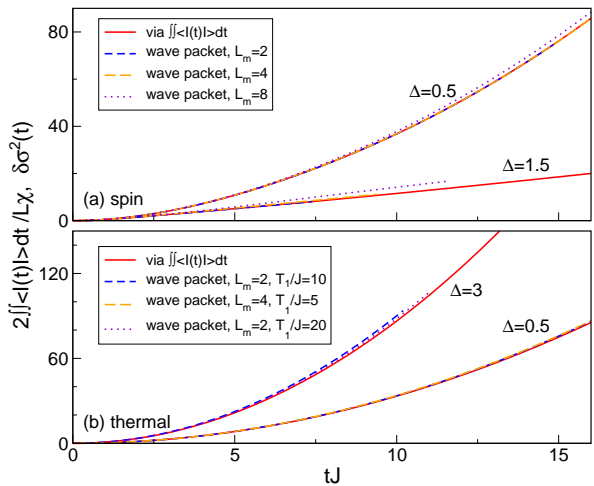}
\caption{(Color online)
The same as in Fig.~\ref{Fig2} (spin and energy quenches for the XXZ chain), but for wave packets of different amplitudes. (a) Spin quench at $T=\infty$. (b) Energy quench with $T_2=\infty$. The amplitude of the wave packet is determined by its size $L_m$ and for the energy quench by the temperatures $T_{1,2}$.
\label{FIGS2}
}
\end{figure}

{\it Details on initial states.--}
Our wave packets are prepared via the following non-equilibrium density matrices $\rho_0$ (that do not commute with $H$). The system is initially cut into three parts,
\begin{equation}
 \rho_0 = \rho_L\otimes \rho_m\otimes\rho_R,
\end{equation}
where a central region of $L_m\ll L$ sites is sandwiched symmetrically between outer (left and right) parts $\rho_{L,R}$ which are in thermal equilibrium. At time $t=0$, the coupling between the three parts is switched on, and the time evolution is governed by the full Hamiltonian.

For energy quenches, the outer and central parts are both prepared in thermal equilibrium, however, at different temperatures $T_1$ and $T_2$, respectively. For spin (charge) quenches in the XXZ (Hubbard) model, the outer parts are prepared in equilibrium at a temperature $T$ while the central sites are fully polarized (empty).

The spatial variance of the (spin, charge, or energy) density distribution $\rho_{s,c,th}$ is defined as
\begin{equation}
\sigma^2_{\nu}(t) = \frac{1}{{\mathcal{N}}_{\nu}}\sum_{l=l_0}^{L-l_0} (l-l_c^\nu)^2\, (\rho_l^{\nu}(t) -\rho_\tn{bg}^{\nu}) 
\end{equation}
where $\rho_\tn{bg}^{\nu}$ is the bulk background density, $l_c^\nu$ is the center of the wave packet, $l_0$ cuts off boundary effects from the left and right ends, and the normalization constant reads
\begin{equation}
\mathcal{N}_\nu =  \sum_{l=l_0}^{L-l_0}(\rho_l^{\nu}(t) -\rho_\tn{bg}^{\nu})~.
\end{equation}
When plotting $\sigma^2$, we always subtract the initial value at $\sigma^2_\nu(t=0)$.

{\it Amplitude dependence.--} The generalized Einstein relations are strictly justified only in the limit of small local perturbations, i.e., wave packets of small amplitude. However, we observe that they still hold even away from this limit. This is exemplified in Fig.~\ref{FIGS2} for the XXZ chain. Fig.~\ref{FIGS2}(a) shows data for spin quenches and various sizes of the central region, illustrating that only slight deviations in the Einstein relation are observed even for a wave packet of $L_m=8$ sites. Note that we prepare the spin-wave packet via fully polarized sites, which is a already a strong perturbation by construction (we have verified that creating only small perturbations using non-fully-polarized sites indeed yields the same results). Figure~\ref{FIGS2}(b) shows the amplitude dependence for energy quenches (i.e., the dependence on the size $L_m$ of the wave packet as well as on the temperatures $T_1$ and $T_2$).

\end{document}